\documentclass[10pt,twocolumn]{article}
\usepackage[utf8,]{inputenc}
\usepackage{amsmath}
\usepackage{amsfonts}
\usepackage{amssymb}
\usepackage{makeidx}
\usepackage{graphicx}
\usepackage{makeidx}
\usepackage{graphicx}
\usepackage{flafter}
\usepackage{placeins}
\usepackage{hyperref}
\usepackage{textcomp}
\usepackage{grffile}
\usepackage{lipsum}
\usepackage{caption}
\usepackage{dblfloatfix}
\usepackage{floatpag}
\usepackage{float}
\usepackage{subfigure}
\usepackage{booktabs}
\usepackage{rotating}
\usepackage{lscape}
\usepackage{rotating}
\usepackage{pdflscape}
\usepackage[table]{xcolor}
\usepackage{background}
\usepackage{dblfloatfix}
\usepackage[a4paper]{geometry}
\usepackage[symbol]{footmisc}

\usepackage[labelfont=bf,textfont=md]{caption}

\usepackage[longnamesfirst, authoryear]{natbib}

\usepackage[nottoc,notlof,notlot]{tocbibind}

\everymath{\displaystyle}

\backgroundsetup{
	position=current page.east,
	angle=-90,
	nodeanchor=east,
	vshift=-5mm,
	opacity=0,
	scale=3,
	contents=Draft
}

\author{Mathew Alex, Relecura Technologies Pvt. Ltd.\\email:\href{mailto:mathew@relecura.com}{mathew@relecura.com}}	
\title{Quantum Technologies: A Review of the Patent Landscape\textsuperscript{*}}
\date{\today}
\begin{document}
	\twocolumn[
	\begin{@twocolumnfalse}
		\maketitle
		\begin{abstract}
			Quantum Technologies is a term that is getting broader with every passing year. Nanotechnology and electronics operate in this realm. With the invention of industry-disrupting algorithms like Shor's algorithm that can break RSA encryption on a quantum computer and Quantum Key Distribution, which offers unconditional security in theory, investment is pouring in. Here we taxonomize and analyze 48,577 patents in this area from 2015 to present captured with a comprehensive query in Relecura's patent database. The author's subject experience, along with the company's AI-based tools and scholarly literature, were utilized to make this highly subjective choice of taxonomy. Though most Patent Landscape Analysis Reports consider a single technology, geography, or company, we have tried to give a holistic overview of these technologies as a whole due to their collaborative and intertwined nature. The physics of each technology and its role in the industry is briefly explained where possible.
		\end{abstract}
	\end{@twocolumnfalse}
	]

	\newpage
	\tableofcontents
    \setlength{\parindent}{5ex}

\footnotetext[1]{Work in progress.}

\section{Introduction}

The extinction of dinosaurs possibly gifted our ancestors a jackpot offer to climb the food chain to dominate the earth. Over the last few centuries, we made unprecedented advances no other species could ever dream of; today, we possess the technology to destroy our planet several times over and cure the largest pandemic ever. History closely followed advancements in science, often shaping ideologies, rather than the other way around. 

The technological advancements were social revolutions of their time. The steam engine and the internal combustion engine made raw mechanical power so cheap that the economy freed humans to work on more creative pursuits. The information revolution accelerated by the transistor's discovery made computers fast and affordable that they enabled us to advance research and industries and to collaborate internationally to create a truly unified global community. Yet, by the end of the last century, we bumped into a few roadblocks that could circumscribe this saga of progress.

To keep up with the exponential pace of innovation, we need faster computers. Dennard scaling (\cite{dennard}) states that shrinking the features of a chip allow running it at a higher clock speed for the same power; in other words, power demand scales with the area while performance scales with the number of transistors. Some ingenious ideas always redeemed the growth of computational power through packing more transistors per area every time in history when it was about to fail. CMOS technology took over when scaling bipolar transistor logic became impractical. When CMOS plateaued, Dennard scaling ended in 2004, and multicore technology met the performance demand (\cite{moore}). As state of the art moves from 7nm to 5nm in 2020, the performance boosts are not stellar, and leakage currents due to quantum tunneling are becoming an extensive hassle to circumvent as we probe deeper. One quantum technology may be the answer to this roadblock that was anticipated decades ago, at least for certain types of computation.

The second roadblock is more profound, thrusting deep into the quantum nature of the universe. The memory requirement of computer simulations of quantum systems exponentially grows with the number of particles (or degrees of freedom), making such systems impossible to model even with supercomputers. Many emergent phenomena like high-temperature superconductivity remain obscure due to their quantum description’s complexity and the impasse with numerical simulations (\cite{manybody}). Some scientists are even moving to machine learning approaches to go beyond the traditional approximation techniques where they failed to illuminate important unexplored problems (\cite{manybodymachinelearning}).

One of the most promising solutions to these problems came from quantum mechanics itself: rather than trying to mitigate quantum effects, we can use them to our advantage. If quantum many-body systems involve such vast amounts of information, why not use them to store this information? Richard Feynman was one of the first scientists to point this out- \textit{“Let the computer itself be built of quantum mechanical elements which obey quantum mechanical laws. ”} (\cite{FeynmanCompute}). A quantum computer meant that we get access to nature’s tremendous storage capacities, her knack for executing physical phenomena in an instant despite the complexity in describing them, and working with the fundamental building blocks of the universe.

A closely related field, which too was coined by Feynman in his lecture "There's Plenty of Room at the Bottom: An Invitation to Enter a New Field of Physics," 1959 (\cite{roomatbottom}) is nanotechnology. It involves manipulating materials at the scale of nanometers, where the quantum nature of matter is apparent. The efficiency and prowess of biological processes come from the molecular machines facilitating them and the nanostructures they use; nanotechnology is a step in this lead's direction (\cite{nanolife}).

Innovation is driven by the synergy between academia, industry, and the economy, and Quantum Technologies are no different. Patents are an excellent indicator of the pace of innovation, and they can be subjected to a wide variety of analyses based on the bibliometric data. Such statistical analyses are invaluable to researchers and policymakers because patents can quantify otherwise difficult to measure phenomena like technological collaboration, the evolution of the technological space, geographical and company-wise predispositions, etc. Increased patenting and the onset of computerization of the field resulted in reports often exploring thousands or even millions of documents. The patent analysis methodology usually varies between reports depending on the context and individual preference (\cite{standard}).

The difficulty with a patent landscape analysis of Quantum Technologies is multifold. Usually, such reports often focus on a single technology or geography. But there is no clean way to cut away quantum communication from quantum cryptography or nanotechnology from semiconductor devices, at least for this set of patents. Moreover, the forefront of R\&D in this field cannot possibly be covered by sticking to a single geography or company. North America is leading research in quantum computing, thanks to the US's tech giants invested in this field and D-Wave in Canada, the first company to sell computers exploiting quantum effects. China is making news frequently for redefining the limits of quantum communication and cryptography. The considerable overlap between some technologies is a hindrance as well that we had to drop some nodes (see Section \ref{overview}). 

In this publication, we try to give a holistic account of Quantum Technologies, starting from analyzing the patent landscape and making connections to scholarly literature as often as needed. We could not find any study incorporating Quantum Technologies as a whole, including Quantum Information, Nanotechnology, Electronics, etc., in patents and scientific research contexts. Yet, some excellent reviews weighing on different techniques have come before, and this publication has benefited greatly from them.

\section{Literature Review}
\cite{pagerank} used a citation-based recursive patent ranking based on PageRank algorithm(\cite{PageBrin}), the first algorithm Google used to rank webpages, to rank patents according to the information flow they contributed to through forward citations. They studied the results for different values of damping factor, $d$, a measure of how much each node benefits from indirect citations, and observed that $d=0.5$ yields best results for patents rather than $0.85$ used by Brin and Page for webpages. A web surfer may go up to, say, ten webpages following hyperlinks, but researchers are not usually interested in patents cited beyond two levels. Hence, this lower value of $d$ is intuitive. Apart from pulling the most relevant patents from the USTPO database using PageRank values, they interpreted patents as a time-evolving complex system in which new technologies emerge as recombination of the old on a mesoscopic scale. Specifically, the paper established that laser/inkjet printer technology emerged from existing sequential printing and static image production technologies by studying the interaction between US Classes through citations over the years.

While we resorted to subject matter expertise to taxonomize the patents, automated patent landscaping is the future, freeing the experts' valuable time for research. The manual process is tedious and expensive, requiring the subject expert to go back and forth between the query and the result multiple times to capture all the valid patents. \cite{deep} proposed an ingenious automated patent landscaping model using deep learning that achieved the state of the art classification performance. The model used Graph Embedding with Self Clustering (GEMSEC) algorithm (\cite{embed}), which preserves clustering while embedding the graph, to encode the metadata, CPC, IPC, and USPC codes in this case. The abstract is handled by a transformer encoder layer (\cite{transformer}) to learn the latent space of a \textit{word2vec} embedding. They concluded that as it is in manual classification, although context-dependent in general, the CPC codes, the most detailed among the three bibliometrics, guarantee better classification performance.

\cite{renewable} conducted an interesting study about patenting in energy technologies that answers the fast-paced innovation in renewable energy in the 2000s despite sustained low funding. To account for this observation, they introduced a non-linear relationship between the cumulative quantities patents $P$, public R\&D $R$, and production levels $C$ as $P=P_0 R^\alpha C^\beta$, where $P_0$, $\alpha$, and $\beta$ are constants. This model accounted for the discrepancy as a combined effect of public R\&D and increasing market.

\cite{dtf} used a neural network to do data-driven technological forecasting for high-tech companies, which is often a difficult decision for companies to make on human instincts alone. Future research and R\&D directions are influenced by a company's present standing and weaknesses in the patent landscape, sensitivity to their competitors, and market forces. The Deep Technology Forecasting (DTF) framework proposed in this work can predict the complex interactions between companies and technologies by quantitatively analyzing their strengths and weaknesses in the patent landscape, identifying competitor companies and their filing trends, and the collaborative relations among technologies.

\cite{review} extensively studied patents in Quantum Technologies, emphasizing the economics, geographies, key players, and their innovation areas. The publication quantitatively argued that China is leading on Quantum Communications while the US remains the leader in technologies related to the manufacture of quantum computers. The State Grid Corporation of China, the world's largest electricity distributor, is taken as an example to illustrate Chinese advances in quantum communication.

A more physics-centered review of Quantum Computing can be found in \cite{QS_2}, although quantum communication, cryptography, and sensing are cut out. The author, also a top scientist in the field, defined the current state of the art as Noisy Intermediate-Scale Quantum (NISQ) computers and argued that physics simulations might be their only immediate use. He encouraged 'optimism tempered with caution' about this technology, and we closely subscribed to his arguments in this work.

This report is a middle ground between the works mentioned in the last two paragraphs, attempting to balance the analysis between patenting trends and physics. We will explore the patents' overall trends in the next section, including an overlap study between first-level nodes and a short discussion of top countries and top assignees. We used a couple of techniques to evaluate the quality of US and Chinese patents; this was necessary since Chinese patents alone account for half of the patents dominating all the statistics.  In Section \ref{toptech}, we centered the discussion the top technologies from the point of view of top assignees in each field. We also provided a short description of emerging technologies identified by Relecura in Section \ref{emerge}.

\section{The Taxonomy}

Some experts classify quantum technologies into two generations(\cite{secondquantum}). The first generation includes transistors, lasers, etc., which work on quantum mechanical principles. The second generation comprises quantum computers, quantum communication, etc., where we actively manipulate the quantum mechanics underlying the universe for our advantage. This entire article is about the second generation of quantum technologies, which were difficult to separate from their predecessors. In a broader sense, even the incandescent bulb work on a quantum phenomenon, the black-body radiation (the problem that gave birth to quantum mechanics).

Quantum technologies is an umbrella term hosting an arsenal of technologies penetrating every field and finding new applications every day. With the advances in big data analysis and machine learning, we can analyze quantum technologies' patent landscape holistically with limited manual labor. Here we examine the $44,394$ patents from this area filed in the last five years, captured using a comprehensive query in \href{https://relecura.com/}{Relecura}'s patent database. We do not claim these patents to be exhaustive, but the tolerances were reasonable. Irrelevant documents that may be present are not enough to sabotage the overall statistics and the following analysis. We avoided brute force subtraction based on keywords to keep the taxonomy receiving and unbiased towards patents and technologies yet to come.

\newgeometry{left=0.1cm, right=0.1cm, top=0.1cm, bottom=0.1cm}

\begin{figure*}
	\thisfloatpagestyle{empty}
	\centering
	\includegraphics[width=0.8\linewidth]{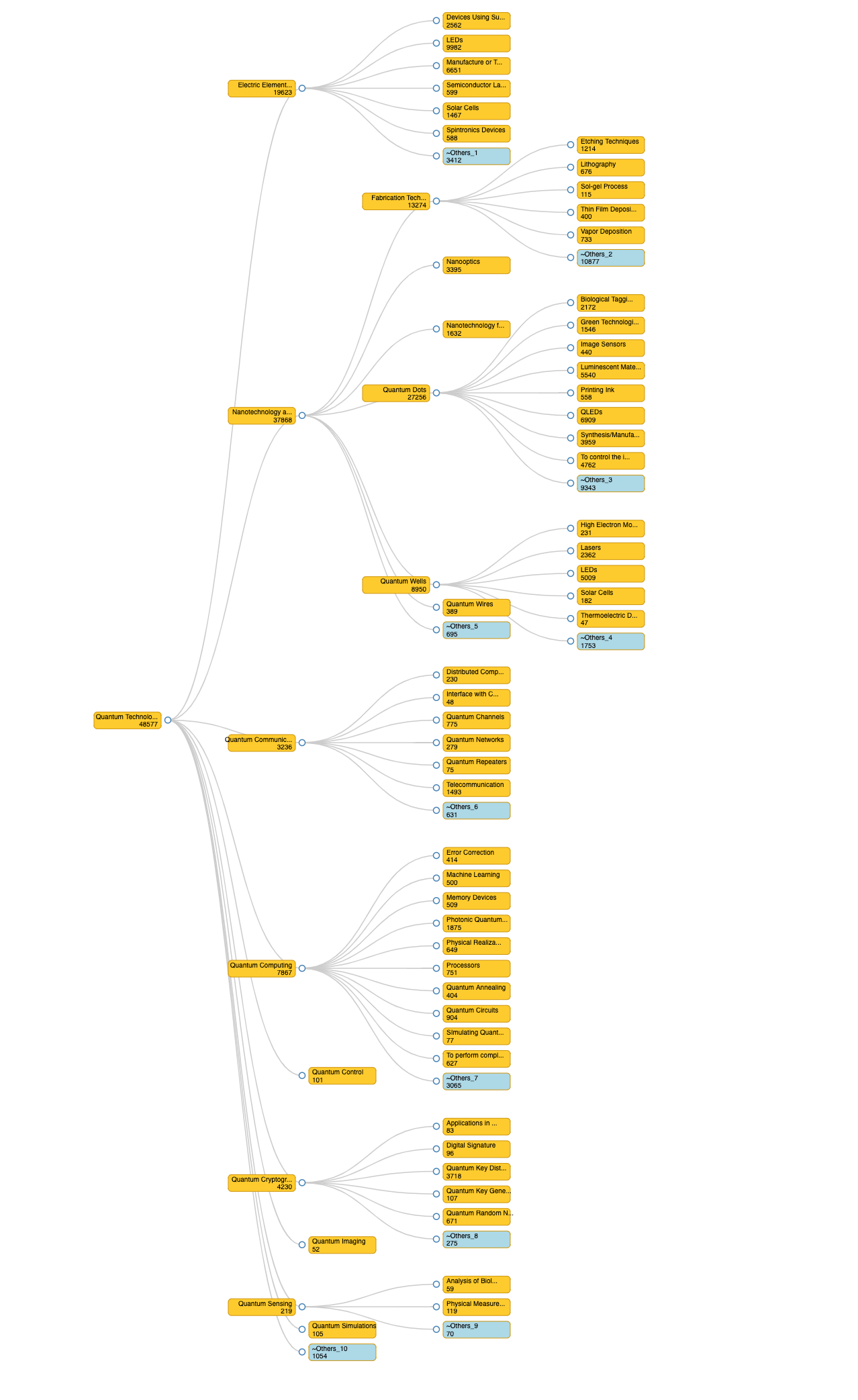}
	\label{fig:taxonomy}
	\caption{Quantum Technologies taxonomy created in Relecura. See this \href{https://t.relecura.com/de8b3d26-651f-11eb-919d-023ada72e4ef}{link} for the up to date interactive web version.}
	\label{Taxonomy}
\end{figure*}
\FloatBarrier
\restoregeometry

\subsection{Overview}
\label{overview}

\begin{figure}[H]	
	\begin{minipage}{0.5\textwidth}
		\centering
    	\includegraphics[width=0.9\linewidth]{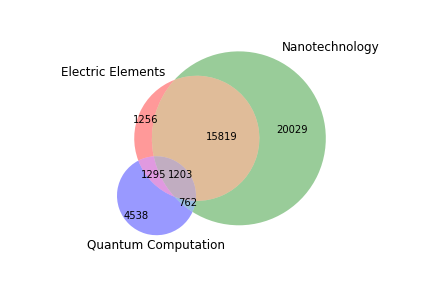}
	\end{minipage}\hfill
	\begin{minipage}{0.5\textwidth}
		\centering
		\includegraphics[width=0.9\linewidth]{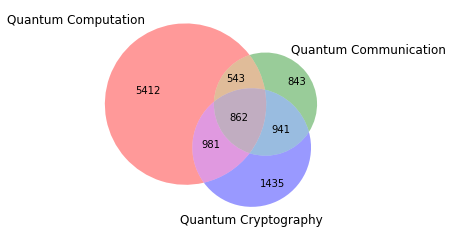}
	\end{minipage}\hfill
	\caption{Overlap between first level nodes}
   	\label{fig:overlap}
\end{figure}

The taxonomy has a significant overlap between first-level nodes (see Figure \ref{fig:overlap}). The \textit{Nanotechnology} and the \textit{Electric Elements} nodes share most patents, and \textit{Quantum Computing},\textit{ Quantum Communication}, and \textit{Quantum Cryptography} nodes have an almost symmetric overlap between each pair. A subject expert will rejoice in the patent sets' overlap reminiscent of the concepts they represent. Modern electronics operate in the nano realm, with new advances in nanotechnology replacing traditional circuit elements. Security is a built-in feature in quantum computation and communication where the collapse of the quantum state on measurement and the no-cloning theorem forbids copying data and eavesdropping (see  \nameref{crypto}). Advancements in quantum communication and cryptography will facilitate the development of the quantum internet, an ideal way to connect quantum processors. 

The organization of this discussion will keep these overlaps in its heart. We will take the \textit{Nanotechnology} node to be representative of  \textit{Electric Elements } as well. The different child nodes are like looking at these patents from a different perspective. The patents in the \textit{Superconducting Devices }child node are unique to \textit{Electric Elements}, and we will discuss them in Section \ref{emerge}. Together with the \textit{Quantum Computation}, \textit{Communication}, and \textit{Cryptography} trio, these are the main patent clusters where innovation is proceeding rapidly. \textit{Quantum Sensing }, too, is an important node, which we will discuss last.

\begin{figure*}
	\centering
	\includegraphics[width=0.6\linewidth]{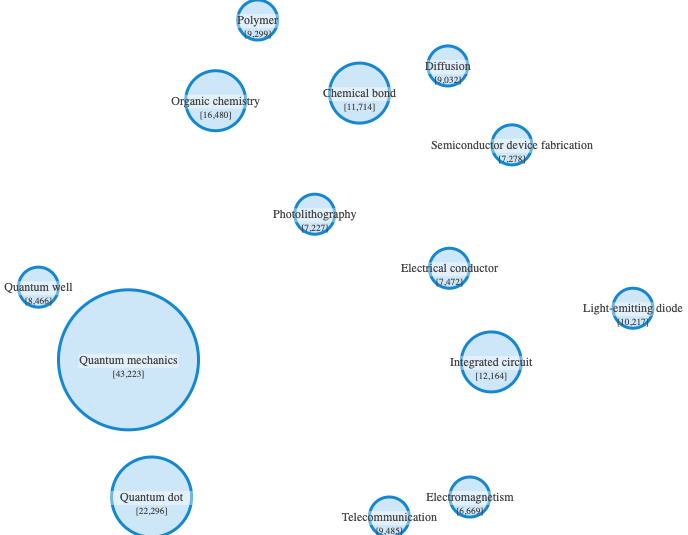}
	\caption{Concept Map}
	\label{fig:concept_map}
\end{figure*}

The concept map (see Figure  \ref{fig:concept_map}) for this patent set generated using Relecura's concept extractor indicates that almost all of the patents utilize quantum mechanics in some form. The largest bubble, \textit{Quantum Mechanics}, is connected with 89\% of the patents, validating this taxonomy.

\FloatBarrier

\subsection{Top Countries}

\href{https://www.reuters.com/article/us-usa-china-patents/in-a-first-china-knocks-u-s-from-top-spot-in-global-patent-race-idUSKBN21P1P9}{China surpassed the US in the number of international patents filed annually in 2019.} The US has been the center of almost all technological revolutions, from the incandescent bulb to the Silicon Valley revolution and the top source of patents since the international system was established. The patent numbers in quantum technologies, when considered in isolation, suggests a Chinese dominance. China holds several times the patents the US has in every first-level node except for \textit{Quantum Computing}, where the difference is only a few hundred patents. \href{https://www.forbes.com/sites/moorinsights/2019/10/10/quantum-usa-vs-quantum-china-the-worlds-most-important-technology-race/?sh=65a3540472de}{China is at the forefront of innovation in quantum communication, and cryptography and the US still hold its edge in quantum computing.}

\begin{figure}[H]
	
	\centering
	\includegraphics[width=0.85\linewidth]{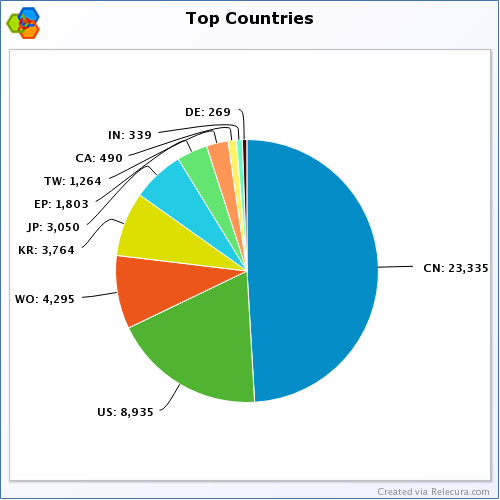}
	\label{fig:parent}
	\caption{}
\end{figure}

\begin{figure*}[htp]
 	\centering
 	\subfigure[]{\includegraphics[scale=0.38]{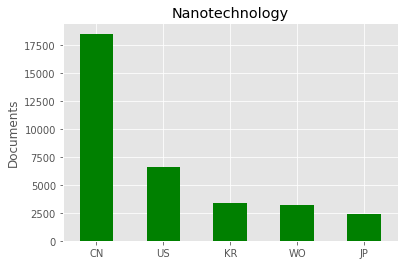}}\quad
 	\subfigure[]{\includegraphics[scale=0.38]{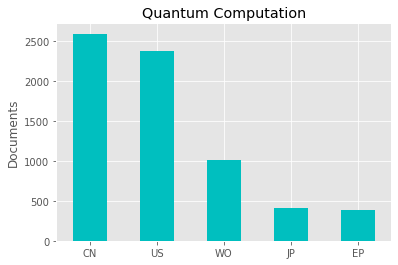}}\quad
 	\subfigure[]{\includegraphics[scale=0.38]{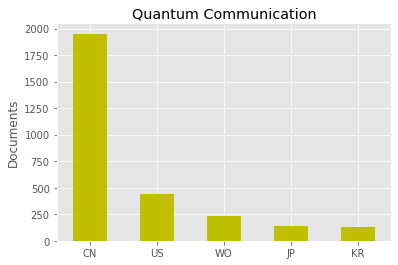}}\quad
 	\subfigure[]{\includegraphics[scale=0.38]{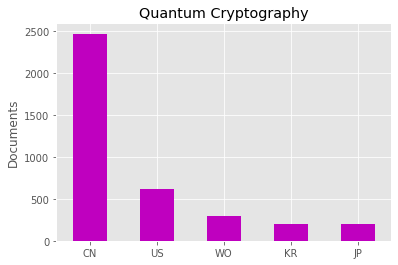}}
 	\caption{Countrywise contribution of patents to major first-level nodes.}
 	 \label{fig:star}
 \end{figure*}

Some resources suggest that most Chinese patents are of inferior quality(\cite{review}) because researchers in China patent more at the expense of novelty to pass the strict objective evaluation of their performance. We used Relecura star rating, a custom rating system that considers key patent parameters like forward and backward citations, geographical and family values, etc., to study this observation. The rating ranges from $0$ to $5$ in increments of $0.5$. Chinese patents caps at $3.5$, whereas the US has patents up to $5$. A normal distribution fit on the star ratings (see  Figure \ref{fig:star} ) arguably reveals that US patents are more valuable. Around 68\% of a normal distribution is concentrated in a standard deviation distance from the mean value. For US patents, it is from 1.37 to 2.81, and in the case of China, it is from 1.13 to 2.27.

One of the crucial factors determining the importance of a patent is the forward citations originating from it. The 8,935 US patents in the taxonomy account for 71,755 forward citations, whereas 23,360 Chinese patents have 44,612 forward citations. A patent citing a previous patent means that the citing patent is building on the knowledge available through the cited patent. 

The citation network provides a representation of the innovation process. Although the primary technique for evaluating this knowledge diffusion, there are concerns over using the citation network, particularly the potential bias or noise added by examiner citations and self-citations that reflects the inventor's knowledge rather than knowledge flow from other patents (\cite{citations}). But even with this noise, the citation data can be a goldmine of information if used with discretion (\cite{citegood}).

\begin{figure*}[htp]
	\centering
	\subfigure[a)US]{\includegraphics[scale=0.36]{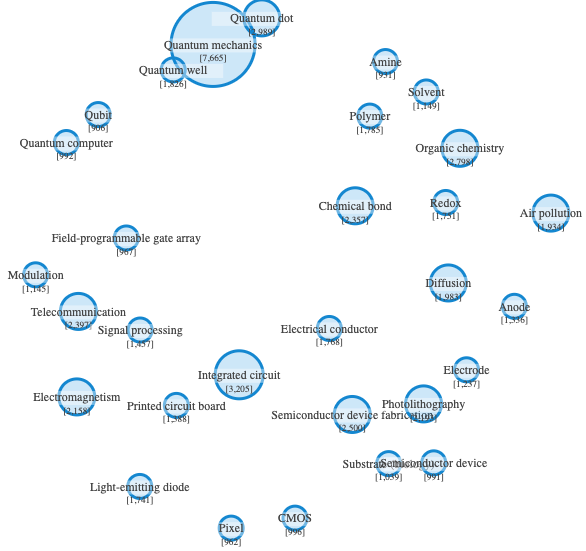}}\quad
	\subfigure[b)CN]{\includegraphics[scale=0.36]{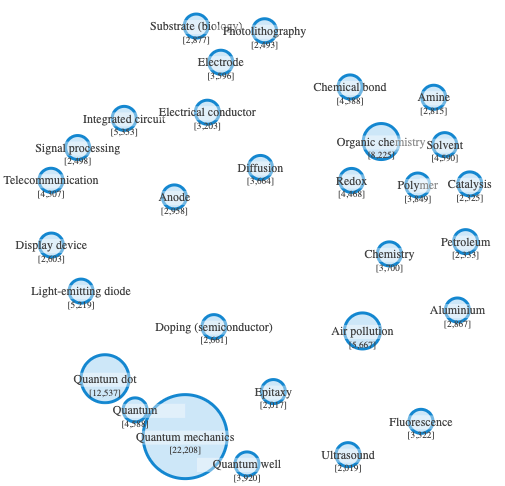}}
	\caption{Concept maps of US and Chinese patents.}
	\label{fig:concept_comparison}
\end{figure*}
\FloatBarrier

We will use the PageRank (PR) algorithm (\cite{pagerank}), which Google used in its initial days to rank websites, to study the citation network. When the internet was a messy place, Larry Page and Sergey Brin came up with this recursive algorithm, which rewards nodes (webpages) that are well-connected to other reliable nodes. It is extensively used to study patents after the algorithm's patent (Patent No. US6285999B1) expired in 2019.  PR is recursively computed for the network using the formula

\begin{equation*}
P_{i}^{(t+1)}=\frac{(1-d) }{N}+d \sum_{j=1}^{n_{i}} \frac{P_{j}^{(t)}}{m_{j}}
\end{equation*}
 
 where $P_{i}
 ^{(t)}$ is the PR value of node $i$ at iteration $t$, $N$ is the total number of the nodes in the network, $n_i$ is the count of incoming links of node$ i$, $m_j$  is the count of outgoing links of node $i$ and $d$ is called the “damping factor”. $d$ controls the contribution to a node through indirect citations and it is set to $0.5$ to study patents rather than $0.85$ used for ranking web pages. This is because on average a web surfer might move through tens of web pages through hyperlinks but the interest in patents does not usually go beyond citations of citations (\cite{pagerank}). We will be using this well tested value that obtained good results in previous studies, setting the iteration convergence tolerance to $10^{-8}$.

\begin{figure*}[htp]
	\centering
	\subfigure[US]{\includegraphics[scale=0.28]{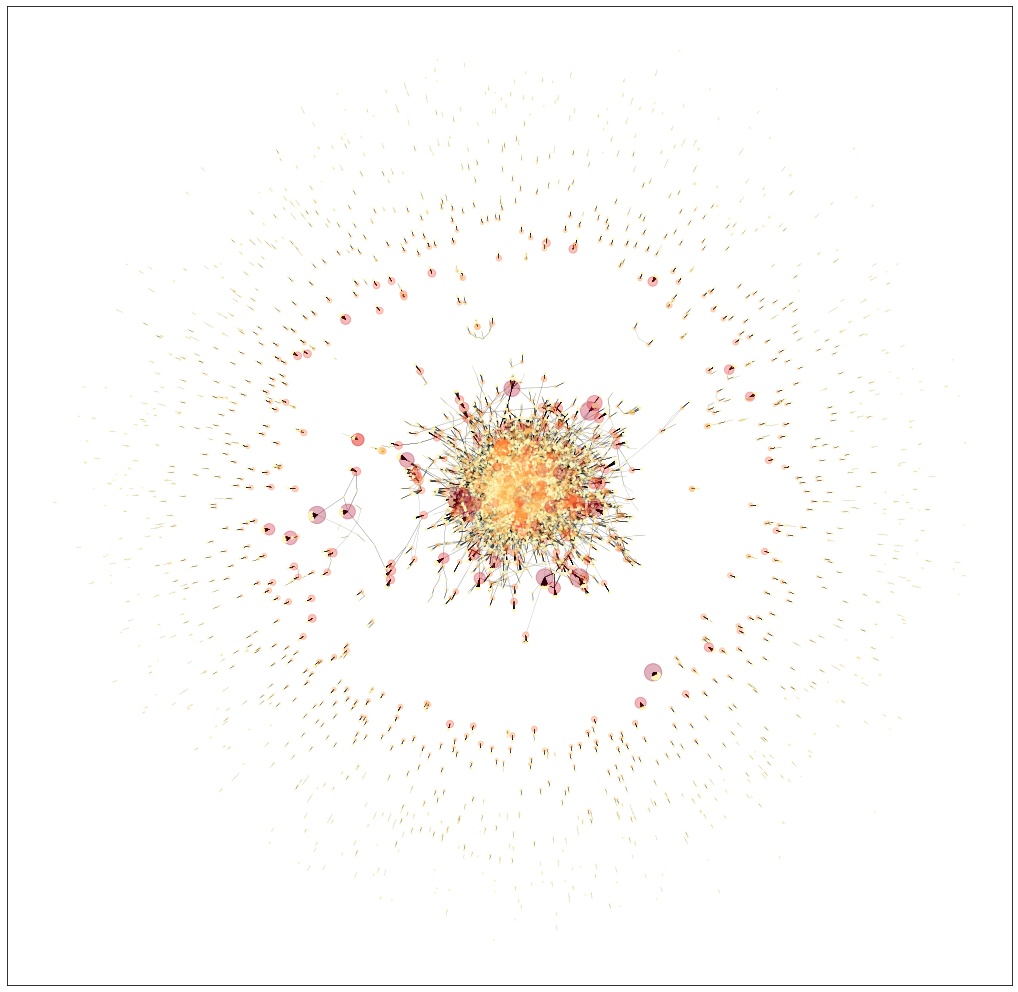}}
	\subfigure[CN]{\includegraphics[scale=0.28]{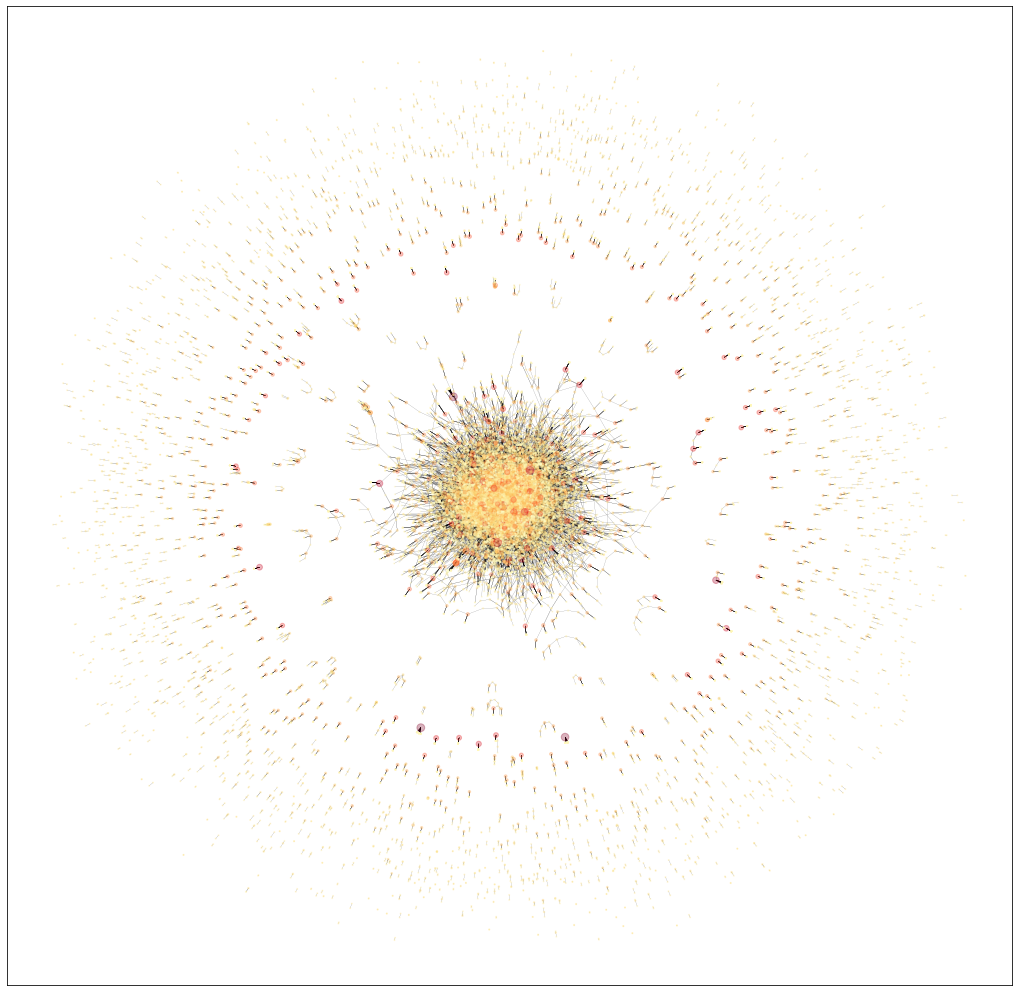}}
   \caption{Patent graph structure illustrating forward citations. In this plot, the node size is proportional to number of connections, and warmer colors mean a higher PageRank score.}
    \label{fig:PR}

\end{figure*}

 We have plotted the network of US and Chinese patents with the nodes color-coded according to PR (see Figure \ref{fig:PR}). Evidently, the US graph form more localized clusters with higher PR, whereas the Chinese graph is more fragmented with lower PR values for the scattered clusters. The drastic difference may be due to the variety of technologies in which patents are filed in China. The concept maps for the US and Chinese patents (see Figure \ref{fig:concept_comparison}) is open for such an interpretation. Chinese patents are more evenly distributed among the top concepts when compared to the US patents. 

The same algorithm is used to pull out the top 20 patents from the taxonomy (see Table \ref{tab:top20patents}), and 18 of them are from the US, and China is absent from the list.

\FloatBarrier

\begin{figure*}[htp]
	\centering
	\subfigure[US]{\includegraphics[scale=0.6]{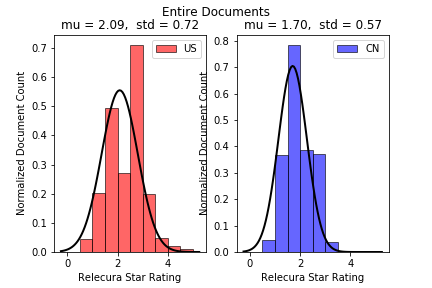}}
	\caption{Star rating distribution of US and Chinese patents fitted with a normal distribution.}
\end{figure*}
\FloatBarrier

 \FloatBarrier

 \begin{landscape}

\backgroundsetup{contents={}}

      \begin{table}[p]
      		\thisfloatpagestyle{empty}
      	\centering
      	\begin{tabular}{cllccc}
      	\toprule
      	No. & Publication Number &                                              Title &  Citations &  PageRank $\times10^8$ & Country Code \\
      	\midrule

      	1  &        US9199842B2 &  Quantum dot films, lighting devices, and light... &                700 &         124623 &           US \\
      	2  &        US9037247B2 &   Non-invasive treatment of bronchial constriction &                301 &         106395 &           US \\
      	3  &        US9049010B2 &  Portable data encryption device with configura... &                297 &         104505 &           US \\
      	4  &        US9000353B2 &  Light absorption and filtering properties of v... &                527 &          98960 &           US \\
      	5  &        US9430078B2 &         Printed force sensor within a touch screen &                291 &          95658 &           US \\
      	6  &        US9019595B2 &  Resonator-enhanced optoelectronic devices and ... &                497 &          89690 &           US \\
      	7  &        US9115348B2 &  Endoribonuclease compositions and methods of u... &                351 &          88545 &           US \\
      	8  &       US10000788B2 &         Rapid and sensitive detection of molecules &                254 &          85708 &           US \\
      	9  &        US9899123B2 &             Nanowires-based transparent conductors &                255 &          84708 &           US \\
      	10 &        US9666702B2 &  Advanced heterojunction devices and methods of... &                342 &          81985 &           US \\
      	11 &       US10572684B2 &  Systems and methods for enforcing centralized ... &                223 &          77716 &           US \\
      	12 &        US8927968B2 &  Accurate control of distance between suspended... &                217 &          75789 &           US \\
      	13 &        US9457139B2 &  Kits for systems and methods using acoustic ra... &                186 &          62477 &           US \\
      	14 &        US9178123B2 &    Light emitting device reflective bank structure &                337 &          59275 &           US \\
      	15 &        US9232618B2 &  Up and down conversion systems for production ... &                178 &          58059 &           US \\
      	16 &        US8932940B2 &  Vertical group III-V nanowires on si, heterost... &                210 &          56184 &           US \\
      	17 &        US9006704B2 &  Magnetic element with improved out-of-plane an... &                248 &          53328 &           US \\
      	18 &     WO2009039854A8 &  MHC multimers in tuberculosis diagnostics, vac... &                145 &          51330 &           WO \\
      	19 &        US9590089B2 &  Variable gate width for gate all-around transi... &                162 &          50451 &           US \\
      	20 &      KR101635835B1 &      Coating method with colloidal graphine oxides &                140 &          49776 &           KR \\
      	\bottomrule
      \end{tabular}

      	\caption{Top 20 patents in the taxonomy according to PageRank with $d=0.5$.}
      	\label{tab:top20patents}
      \end{table}

 \end{landscape}

\subsection{Top Assignees}
The top assignees in the entire patent set are well-known companies. Most of them are here in this category because of the variety in their research. There are four US companies (Alphabet, IBM Intel, and Northrop Grumman), three Chinese companies (BOE Technology, Chinese Academy of Sciences and TCL Corporation), two Korean companies (Samsung and LG), and a single Japanese company (Toshiba) in the top 10 assignees.

\begin{figure}[H]
	
	\centering
	\includegraphics[width=0.85\linewidth]{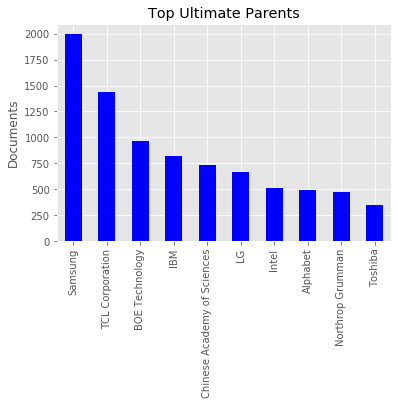}
	
	\caption{}
\end{figure}

The US companies are almost centered on the \textit{Quantum Computing }node with an overflow to the \textit{Nanotechnology} node from their investment in the semiconductor industry. 435  of Alphabet's 505 patents and 700 of IBM's 876 patents go to \textit{Quantum Computing}, which is not a surprise as they are the industry leaders of the field. Intel has got 459 patents in \textit{Nanotechnology} and 280 in \textit{Quantum Computing}; they are working on building quantum processors collaborating with \href{https://qutech.nl/}{QuTech} in the Netherlands. Northrop Grumman, one of the world's largest defense technology providers and maker of the renowned B-2 stealth bomber, is actively exploring new disruptive quantum technologies. Their portfolio is concentrated in \textit{Quantum Computing} (283) and \textit{Superconducting Devices} (388 patents).

TCL Corporation (the makers of  Blackberry phones from 2016 to 2020), which manufactures mobiles display panels etc., is patenting heavily in  \textit{QLEDs} (704 patents) and \textit{Fabrication Technologies} (528 patents). BOE Technology, one of the world's largest producer of LCD, OLED, and flexible displays, has a similar patent profile (434 patents in \textit{Fabrication Technologies} and 570 patents in \textit{QLEDs}). Chinese Academy of Science, the only academic institution in the top 10 assignees, has a well-balanced portfolio in quantum technologies with patents in all first-level nodes.

LG is more swayed towards \textit{Semiconducting Devices} and \textit{Nanotechnology}, which is expected since its primary trade is consumer electronics. The other Korean giant in top assignees, Samsung, shows similar trend except that the numbers must be scaled.  Almost all of its patents come under \textit{Nanotechnology}.
	
Though Toshiba is known for its consumer electronics products, \href{https://www.reuters.com/article/us-toshiba-cyber/toshiba-targets-3-billion-revenue-in-quantum-cryptography-by-2030-idUSKBN2730KW}{ it has diversified its business model to include IT solutions like quantum cryptography in 2020.} They hold hundred plus patents in \textit{Quantum Computing, Communication, and Cryptography}.

\section{Top Technologies}
\label{toptech}
This report cannot possibly cover every leaf level node. We have established that \textit{Nanotechnology and  Quantum Computing,  Communication, and Cryptography} are the major first-level nodes in this taxonomy. \textit{Nanotechnology} or even its child node \textit{Quantum Dots} taken in isolation, account for more than half of the entire patent set. We follow a weighted approach in our analysis, with these nodes taking the most precedence and detailed discussion.

\subsection{Nanotechnology}

The most populated and dynamic node in this taxonomy is \textit{Nanotechnology}, and the low-dimensional nanostructures (\textit{Quantum Dots}, \textit{Wells}, and \textit{Wires}) dominate patents in this node. In a way, this node captures the spirit of quantum technologies.

The development of semiconductor technologies in the last century was mostly limited by the difficulty in synthesizing materials with the desired energy levels. When you choose a material for its energy levels, you have to live with its chemistry. LEDs were a revolution in lighting technologies with their power efficiency and long life span, but this innovation was bottlenecked by the delay in making a blue LED. The first blue-violet LED was developed using magnesium-doped gallium nitride at Stanford University in 1972. Isamu Akasaki, Hiroshi Amano, and Shuji Nakamura were awarded the Nobel Prize in Physics in 2014 for the  invention of  the blue LED by growing high-quality Gallium Nitride crystals and creating p-type channels in it.

\begin{figure}[H]
	\centering
	\includegraphics[width=0.9\linewidth]{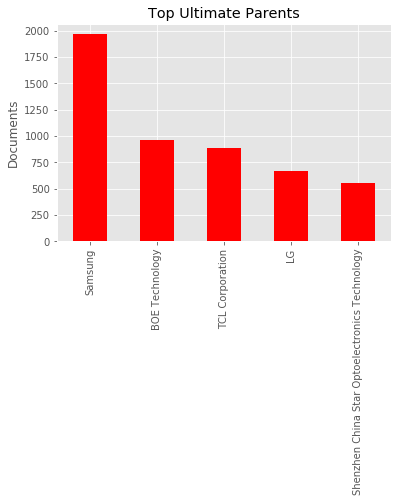}
	\caption{Top Assignees in Nanotechnology}
\end{figure}
\FloatBarrier

The addition of blue LED completed the color spectrum with the already available red and green LEDs, and this trio made its way into LED screens and the lighting industry (thank blue LEDs you can read this article). The availability of high wavelength (and energy) blue LEDs meant that other colors could be generated by phosphors or combining them with red and green LEDs. This invention lightened the power grid load and expedited solar energy adoption because of its power efficiency. But it took decades to go from blue-violet LED to a pure blue LED.

Nanotechnology, the engineering of functional systems at a molecular scale, offers a different approach to this difficulty of synthesizing crystals for their energy levels. The electrons in a crystal see a practically infinite periodic lattice, and the solution of the Schrodinger equation (the differential equation that governs quantum behavior) for this system is energy bands typical semiconductors. However, when the charge carriers are confined in one or more dimensions to the order of their de Broglie wavelength, the scale at which quantum effects dominate, they will know that the periodicity is gone, and the bandgap changes.

Confinement doesn’t always mean that the material is of the length scale of carrier confinement. Quantum wells (2-d nanomaterials) are grown as alternate layers of semiconductors with different bandgaps. The progress of this field was entirely due to the developments in advanced crystal growth techniques like molecular beam epitaxy and microscopy techniques like Scanning Tunneling Microscope, which provided unprecedented visualization of atoms and bonds, and as \href{https://www.ibm.com/ibm/history/exhibits/vintage/vintage_4506VV1003.html}{demonstrated in 1990}, capable of moving individual atoms around (see \href{https://www.youtube.com/watch?v=oSCX78-8-q0&ab_channel=IBM}{A Boy and His Atom} to know how proud IBM is of this technology).

While research in the past took the quantum energy levels as absolute, nanotechnology goes to the next level to make novel materials with tunable electrical, optical, and physical properties. Devices crafted like this are aggressively replacing traditional circuit elements with their futuristic efficiency and speed.

The low dimensional nanostructures are an exciting area of research working at the intersection of physics and chemistry. They operate at the mesoscopic scale, between microscopic and macroscopic laws of physics, larger than a few electrons or molecules but small enough that their degrees of freedom must be treated fully quantum mechanically. Scientists are dealing with hardcore quantum mechanics with all its quirks here. The research in this field has even unveiled many things about the movement of electrons in traditional devices.

\subsubsection{Quantum Dots}
\label{qd}
Low dimensional nanostructures (32,529 patents), particularly quantum dots (25,058 patents), absolutely crushes everything else in sheer numbers in our taxonomy. Since it accounts for more than half of the patents, we will go down this node, most representative of nanotechnology and quantum technologies in general, to discuss its applications.

Lower dimensional materials are materials restricted to atomic scales in one or more dimensions (\cite{lowdim}): graphene is two-dimensional, and a quantum dot is zero-dimensional by this definition. Confinement of charge carriers in such structures leads to exotic electronic and optical properties tunable by their size and composition. For example, quantum dots (QD) can be adjusted to fluoresce from blue to red by increasing the particle size. They find potential applications in LEDs, displays, lasers, and biological imaging. We will showcase a  few important applications of QDs here.

\begin{enumerate}
	\item \textbf{Biological Tagging and Labelling}
	In biological tagging, QDs have significant advantages over conventional fluorophores (a fluorescent tag that selectively binds to a specific region or functional group on the target molecule). The Green Fluorescent Protein (GFP), naturally occurring in a particular jellyfish, and other chemically synthetic fluorescent dyes were the most common biological tagging methods. QDs shine here due to their efficiency in capturing light to give a brighter image for the same irradiation compared to traditional fluorophores. Thanks to their inorganic origin, they don’t undergo photochemical degradation from prolonged exposure to the excitation source. Their size-tunable nature allows a complete gamut of colors. QDs find use in drug discovery, single protein tracking, and disease detection. Maybe someday, we will have the ability to light up any biological phenomena in any color scheme!
	
	Since this is still an experimental technology, most assignees are academic institutions: University of Jinan (31 patents), Duke University (23 patents), Harvard (23 patents), US National Institute of Health (21 patents), and French National Centre for Scientific Research (18 documents) are among the top 10 assignees. Nanoco Technologies (41 patents) and Shenzhen Fortense (31 patents) are the top two assignees.
	
	\item \textbf{Image Sensors}
	QDs make excellent photodetectors, and QD image sensors are on the way. Traditional CMOS sensors capture the electrons ejected from the semiconductor pixels through metal contacts and traces, which reflect part of the light, decreasing efficiency and contributing to poor low light performance. The solution to this is the back-illuminated sensor,  with readout electronics under the detector, but it costs way more than its predecessor. 
	
	Smartphone cameras have developed so fast in the past few years that DSLRs only have an edge under challenging shots that do not matter to most consumers, but the software has been doing most of the heavy lifting. Google thrived on computational photography to establish the Pixel lineup launched in 2016; their software was so good that they haven’t upgraded the image sensor (Sony IMX 363) for their last three generations of phones. But all of its competitors caught up in 2020 that the differences between brands are subtle and computational photography has plateaued.
	
	QD image sensors have the potential to take the industry from here. A QD sensor is fabricated in the same way except for the silicon photodetector part; instead, QDs are suspended in a solution and sprayed onto the wafer, reducing the cost. This setup offers the benefit of a back-illuminated sensor without its fabrication cost. The photoelectron can hop from one QD to another to reach the nearby electrode. A  thin layer of QD is enough because it can absorb light much better than silicon. They function well in low and high light scenarios offering a better dynamic range(\cite{QD_image}). Due to their size-tunable nature, QDs can be used in infrared cameras as well. Since silicon’s bandgap is higher than the energy of infrared light, present IR cameras have to go through the added hassle of integrating a different semiconductor into the silicon wafer. QD sensor is easy to manufacture for both IR and visible range applications.
	
	Samsung (67 patents), Canon (60 patents), and Fujitsu (26 patents), all of whom manufacture cameras in one form or another, are the top three assignees in QD image sensors. The third top assignee is tying with Fujitsu, InVisage Technologies, founded by Ted Sargent, who pioneered QuantumFilm technology when he was a Professor at  the University of Toronto. QuantumFilm is the suspended version of QDs to be coated on the image sensor mentioned previously. The industry was heated up when Apple, a lowkey player in this area but the company which nailed the camera performance in every iteration of their phones, acquired InVisage. Although Apple is silent about its R\&D, an iPhone with a QD image sensor is likely to release in the near future to claim back the camera industry for hardware.
	
	\item \textbf{QLEDs}
	
	Although quantum dots are an emerging technology, the industry deemed it ripe for commercial displays. Several companies like Samsung are aggressively marketing QLED TVs as the next big thing. Maybe, this might be the only piece of second-generation quantum technology that you can buy right now.

	Out of the 3,138 patents listed under QLEDs, Samsung dominates with 354 patents, closely followed by BOE Technology (307 patents), Shenzhen China Star Optoelectronics (192 patents), TCL Corporation (121 patents). HiSense (58 patents) and TCL producing QLED TVs to compete against Samsung. Even LG (104 patents), who denounces QLEDs in their \href{https://www.lg.com/us/experience-tvs/oled-tv/oled-vs-qled}{website}, labeling it as a gimmicky LCD technology, is actively doing R\&D in this area.

	Even though both OLED and QLED are improvements over the LCD technology, it is hard to crown one of them better. OLEDs can produce light of their to deliver practically infinite contrast and an almost 180-degree viewing angle. They can be manufactured as transparent or flexible displays (the reason for the recent onset of foldable phones and bezel-less phones that tuck the display under the visible screen to attach the controls). However, OLEDs are expensive to manufacture as large panels. The cost will increase as we run out of rare earth elements required, and the screens’ longevity is a serious issue.

	QLED displays are cheaper than the OLED ones, and they are more power-efficient, lasts like LCDs, and offer more color purity and a wider color gamut. While QLED is a fantastic comeback of the backlit technology and offers healthy competition against OLEDs in the TV market, they are unlikely to make their way into smartphone screens due to the availability of flexible OLED screens and the added weight of the backlighting setup. Yet, these corporations’ serious R\&D investment suggests that they can evolve into the next big thing in time.
	
\end{enumerate}	
	
	However, the cost of manufacture is a significant hindrance to the early adoption of this technology, and the toxicity of chemicals involved (for example, Cadmium) is another hit. The EU has issued a temporary relaxation of RoHS (Restriction of Hazardous Materials) for quantum dots, which will expire soon. Biological applications are limited from mass adoption due to the toxicity factor. An actively researched solution is to design QDs to evade breaking down by immune cells to avoid long term retention of heavy metals in the body. Either way a non-Cadmium based QD is urgent (\cite{noncad}). QDs can oxidize in the air, which is not a problem in enclosed display panels, but must be taken into account when designing image sensors. In the meantime, accelerated research in quantum dots is breaking these barriers and finding them more niche uses.

\subsection{Quantum Computing}

This node might be the most magical realm in all quantum technologies with paradigm-shifting implications if ever realized commercially. Quantum computing uses the universe's e fundamental building blocks (single atoms, electrons, protons, etc.) as computational elements. However, they cannot execute irreversible gates like AND and NOT, and they are never meant to replace consumer computers, even though they are Turing complete. So far, only a few algorithms are available that can manipulate quantum mechanics to achieve exponential speedup over classical computers. A quantum computer employing Shor's algorithm can factor prime numbers used in RSA cryptography (RSA numbers), while it is virtually impossible for a classical computer in reasonable timescales. Grover's algorithm can complete an unstructured search in a database of $N$ elements in a time propotional to $\sqrt{N}$. It will be an awful waste of resources to code a quantum computer to play a movie.

\begin{figure}[H]
	\centering
	\includegraphics[width=0.9\linewidth]{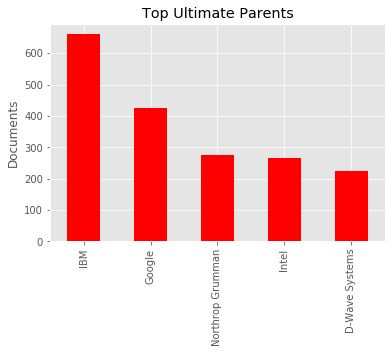}
	\caption{Top Assignees in Quantum Computing}
\end{figure}
\FloatBarrier

Quantum computers ingeniously use quantum superposition, the property of quantum systems to occupy multiple states weighted with different probabilities, and entanglement, the ability of spacially separated entangled particles to act on each other, to parallel process vast amounts of data. An $N$ qubit array can store $2^N$ numbers and parallel process them efficiently(\cite{QC_1}). Quantum computers can evaluate functions parallelly by encoding the inputs as a superposition state and evaluating once. However, we can not assess all of the outputs as one measurement will collapse the state. Even though the information is there, we need to come with algorithms respecting the quantum weirdness to take advantage of it.

Quantum computers work best in problems that are easy for a classical computer to verify an answer's truth, like factoring a composite number. The Shor's algorithm on a quantum computer offers an exponential speed up to the factorization problem. The RSA cryptography system, which relies on the difficulty of finding prime factors of a composite number, is at stake if this algorithm scales from its record of factoring the number 35 (\cite{shor_max}). A quantum computer employing a few thousand qubits can break the RSA and Elliptic-Curve Cryptography (ECC) systems.

Niche algorithms have been developed for other mathematical problems as well that quantum computers changed the paradigms of classical computing; they even introduced a new complexity class (see \href{https://en.wikipedia.org/wiki/BQP}{BQP}) in the theory of computation. Even though scientists came up with revolutionary algorithms decades ago, they have not materialized on a real quantum computer to have any physical impact as of yet. The accelerated research in recent times is trying to mitigate this gap. The most number of qubits in a quantum computer is less than 100 as of now, and this must shoot up to the hundreds and the thousands for quantum algorithms to change the world. But still, they do not pose a challenge to personal computers because only a handful of algorithms are available, offering quantum speedup in very specific problems. Quantum computers are likely to reach out to the world through the cloud as most of the current realizations operate at cryogenic temperatures, requiring bulky cooling systems.

The best in the business is taking up this challenge. Even though the world's largest cloud service provider by market share, Amazon, does not have a strong presence in this taxonomy, they are well placed in the business performing an integrating role between different quantum processors developed by other companies. \href{https://aws.amazon.com/braket/}{Amazon Bracket}, released last year, allows customers to build and simulate quantum algorithms on simulators as well as on different quantum computers. They offer D-Wave's quantum annealing processors (see Section \ref{annealing}) and gate-based computers from Rigetti and IonQ.

The big names in the computing industry here competing against each other, often collaborating with smaller companies, to be the leader in this technology of the future. Out of the 7,867 patents, the patent champ IBM holds 700 patents to lead in this area. On their shift to research and consulting services, they kept close to quantum computing to create the best python library \href{https://qiskit.org/}{Qiskit}, which is widely used in academia, to simulate quantum circuits and execute them on IBM's quantum processors through the cloud. Google offers \href{https://ai.googleblog.com/2018/07/announcing-cirq-open-source-framework.html}{Circ-q}, and Microsoft offers \href{https://docs.microsoft.com/en-us/quantum/}{Q\#} as competition, but none of them provide access to real quantum computers. Plus, IBM collaborates with fortune 500 companies and top universities through the IBM Quantum Network program for accelerated research and early commercialization of quantum computing solutions. The next contenders are  Google (434 patents), Northrop Grumman (282 patents), Intel (280 patents), and D-Wave Systems (231 patents). 

Most of these top assignees work to achieve quantum supremacy, i.e., to demonstrate that a quantum computer can solve a classically intractable problem. Google claimed in October 2019 that it \href{https://ai.googleblog.com/2019/10/quantum-supremacy-using-programmable.html}{achieved quantum supremacy using a new 54-qubit processor named “Sycamore”}(\cite{supremacy}). However, this is disputed \href{https://www.ibm.com/blogs/research/2019/10/on-quantum-supremacy/}{by IBM} and \href{https://www.nature.com/articles/d41586-020-03434-7}{other researchers in the field.} Even though a milestone in computing, quantum supremacy has a relaxed definition: the algorithm does not need to perform anything useful, which means that it can be achieved before significant advances in error correction. It is likely to be achieved soon, if not already.

Another specialty of this area in the patent landscape is the presence of universities. Along with the big names like MIT, Caltech, etc., Chinese and Korean institutions are filing a comparable number of patents to industries. \href{https://trumpwhitehouse.archives.gov/articles/trump-administration-investing-1-billion-research-institutes-advance-industries-future/}{The US is increasing funding for quantum computing research at universities} and the academic contribution is expected to grow.

\subsubsection{Quantum Annealing} 
\label{annealing} 
 'Annealing' means to heat a material past its recrystallization temperature and cool it to remove defects and internal stresses. Nature loves to settle to the minimum energy configuration if given time, no matter how complicated the optimization problem is in the number of variables involved; this is how gems, rather than coal, form under slow geological processes. 

Optimization techniques look at a function's local geometry to move to a better solution, but they are prone to get trapped at local extreme values. Stochastic optimization methods add an element of randomness to free the algorithm from being trapped at such values.  Classical/simulated annealing is a stochastic optimization method in which these probabilistic jumps mimic the statistical physics description of the annealing process. Quantum annealing improves the classical annealing technique by replacing the simulated 'thermal fluctuations' with quantum fluctuations in real quantum systems.  

Quantum annealing uses a quantum mechanical system in which the cost function acts as potential energy, a term nature tries to minimize. Whereas the classical algorithm had to climb the height of the function's 'hills' to find the next minimum, the quantum system can tunnel through the barrier as the width (not height) gets small. Hence this algorithm outperforms classical annealing, at least in theory, if the function landscape has many high but thin hills surrounding the minima.

There is a company that does this and only this. Rather than putting out a universal quantum computer that is Turing complete (capable of computing everything computable and able to run all algorithms like the Shor's algorithm), they capitalize on quantum annealing processors, which only do quantum annealing. It is D-Wave Systems, a Canadian company, that sold the world's first computer exploiting quantum effects. They followed the industry canon to put out a first-generation product in the market than to thrive on claims. D-Wave might be the best example for a company that successfully capitalized on quantum technology. Soon after its inception at the University of British Columbia, its customer base grew to include Lockheed Martin, Google, NASA, and Los Alamos National Lab. \href{https://www.dwavesys.com/applications}{Their annealing computers are finding more applications every day in optimization, machine learning, and material science.}

Quantum annealing is that important, and the promises of speed up are groundbreaking. D-Wave's latest release, Pegasus, has 5,000 qubits, but this is not a significant feat in error correction (see next section) since they are designed to do annealing alone. We could not find any conclusive studies on the D-Wave processor's performance.

\subsubsection{Error Correction}

Decoherence is the Achilles heel of quantum computing, confining it to prolonged infancy. Completely isolating quantum systems, qubits in this case, from the rest of the universe is next to impossible; anything can couple with the system's dynamics to send the quantum state into the unknown, sabotaging the calculation. Efficient quantum algorithms make use of large scale interference between qubits, which is very delicate. It is difficult to achieve this coordination between qubits while protecting them from unwanted influences. This problem was once thought to be so forbidding that quantum computing would never work. Error correction in classical computing relies on redundancy, but that is not an option in quantum computing due to the No Cloning Theorem, which states that no algorithm can copy arbitrary unknown quantum states.

Multi qubit error correction is so daunting that many introductory reviews shy away from them. However, the mathematics of quantum mechanics has led to many elegant formulations, including the stabilizer codes (\cite{stabilizer}). Single qubit error correction relies on coding a logical bit in a superposition of multi-qubit states. For example, Shor's code encodes one logical bit into a nine qubit state, and subsequent algorithms reduced the number from nine.

Google (38 patents), IBM (32 patents), and Microsoft (24 patents), who are also the industry leaders in quantum computers, are the top assignees in quantum error correction. IBM is expecting to operate a 1000+ qubit quantum computer by 2023, according to their roadmap. At this scale, quantum computers can change the world and move away from the textbook example scale problems it solved in its infancy. Better error correction codes are all there is between the 65 qubit and the 1,121 qubit computers mentioned in \href{https://www.ibm.com/blogs/research/2020/09/ibm-quantum-roadmap/}{IBM's roadmap}.

\subsection{Quantum Cryptography}
\label{crypto}
All present cryptographic systems rely on hard one-way mathematical problems; for RSA cryptography, it is the integer factorization, and for Elliptic Curve Cryptography (ECC), it is the discrete logarithm. When quantum computers catch up, the newly found computational power will take the ‘hard’ label away from these problems, so post-quantum cryptography is a primary concern for industries as well as governments. If quantum computing develops as planned, it will break both RSA and ECC. However, these two cryptography systems are challenged by Shor’s algorithm, and other unchallenged classical cryptographies that can resist attacks from a quantum computer are already available. On the other hand, there is a quantum solution to this quantum problem. 

\begin{figure}[H]
	\centering
	\includegraphics[width=0.9\linewidth]{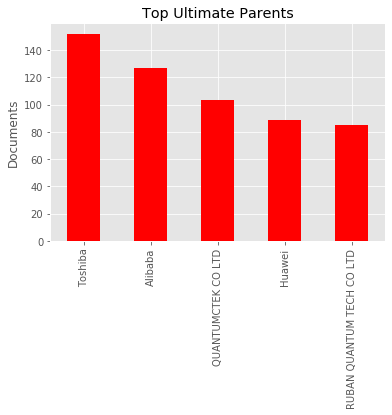}
	\caption{Top Assignees in Quantum Cryptography}
	
\end{figure}
\FloatBarrier

\subsubsection{Quantum Key Distribution}

QKD eliminates the vulnerability in transmitting encryption keys when shared through classical communication channels. Whereas pre-quantum cryptography relies on its mathematical complexity to ensure security, QKD relies on the no-cloning theorem, our adversary in quantum error correction. QKD dominates the quantum cryptography node with 3,718 patents. 

QKD provides unconditional security, the holy grail of cryptography, in the exchange of encryption keys. Relying on hard one-way problems in mathematics for encryption has the problem that it gets vulnerable as computational power catches up with time. Would anyone purchase a lock that gets weak over time? It depends. For example, the health records of a person must remain private in his lifetime. If health data is compromised, insurance companies, employers, etc., can make discriminating decisions to increase revenue. Anyone can record today's internet traffic and the public keys to decrypt them later, possibly before the data loses its relevance considering the growth of ordinary computers, let alone quantum computers. 

The public key shared using QKD (the details are omitted here for brevity, see this excellent review for details (\cite{qkd})) cannot be intercepted by an eavesdropper, no matter how equipped he is. The measurement act will change the message carrier's quantum state to let the receiver know that someone tried to intercept the transmission. 

R\&D in this area focused on China, with this country holding the most patents (2,465 patents), way ahead of the US (445 patents). China has recently been in the news for pushing the limits of QKD to establish a secure communication link to the low earth orbit satellite Micius over a range of 1,200 kilometers(\cite{satellite}). Terrestrial communication links are limited to a few hundred kilometers due to transmission losses in optical fiber; the no-cloning theorem forbids noiseless signal amplification. Free space communication has the advantage that the photons have to get through roughly 10 kilometers of the atmosphere with negligible absorption. China's feat takes us closer to the dream of a global secure quantum network. Though channel loss in optical fiber is a hindrance, quantum repeaters can solve this problem, but this technology is still immature for practical implementation. Chinese scientists have established optical fiber links of up to 404 kilometers as of now (\cite{opticalfiber}). 

Although four of the top 5 assignees in this area are Chinese, Toshiba, a Japanese company, leads this group. The company is already \href{https://www.toshiba.co.jp/qkd/en/products.htm}{offering QKD systems} to address the data breach problems that cost a fortune to companies and government organizations. IBM \href{https://www.ibm.com/security/data-breach}{estimated} that the average cost of a data breach is 3.86 million USD, and it takes an average of 280 days to identify and contain the breach. Toshiba's solutions are appealing in this context of the already \href{https://www.informationisbeautiful.net/visualizations/worlds-biggest-data-breaches-hacks/}{shaken communication infrastructure.} They have been involved with QKD from the 90s onwards after starting the Toshiba Research Laboratory in Cambridge (1991). The company plans to own a fair share of the multibillion-dollar market they expect for QKD in the 2030s.

The major US companies and government agencies, which are actively filing patents in quantum computing, have a small number of patents in \textit{Quantum Communication} and \textit{Cryptography}. This might be correlated to the government level policy about QKD; \href{https://www.nsa.gov/what-we-do/cybersecurity/quantum-key-distribution-qkd-and-quantum-cryptography-qc/}{the NSA is vocal about the shortcomings of QKD as post-quantum cryptography}. Even though the eavesdropper cannot de-code the message, he can disrupt the communication by reading the information passed. Plus, the NSA argues that the infrastructure cost is too much, and the theoretically secure QKD is known to be vulnerable in its physical implementations (\cite{qkdattack_1}\cite{qkdattack_2}\cite{qkdattack_3}).  Research in the attacking side is coming up with new strategies to eavesdrop or interupt the communication, and the defending side is trying to eliminate these vulnerabilities (\cite{qkdprotect_1}\cite{qkdprotect_2}).

 \href{https://csrc.nist.gov/projects/post-quantum-cryptography}{NIST is actively looking at post-quantum cryptography proposals}; early adoption of such an algorithm would ensure that the data encrypted will not be breached while it is sensitive. Using a symmetric key cryptographic system like \href{https://en.wikipedia.org/wiki/Advanced_Encryption_Standard}{Advanced Encryption Standard (AES)} (champion of the last NIST contest in 2000 and US national standard since then)with a doubled key size would be a good idea in the meantime. Symmetric key systems are vulnerable to Grover's algorithm, which offers a square root speed up in a brute force attack, and a 256-bit key can provide the same security as a 128-bit key in the pre-quantum world (\cite{postquantum_1}). 

QKD requires specialized hardware, and the industry is analyzing the price to performance ratio. Existing cryptography systems are so optimized over time, and some of them even made into instruction sets of Intel and AMD processors.  QKD is likely to replace high-stakes communications at first, considering the implementation costs and bandwidth limitations in its present form.

\section{Emerging Technologies}
\label{emerge}

\subsection{Superconducting Devices}
\begin{figure}[H]
	\centering
	\includegraphics[width=0.9\linewidth]{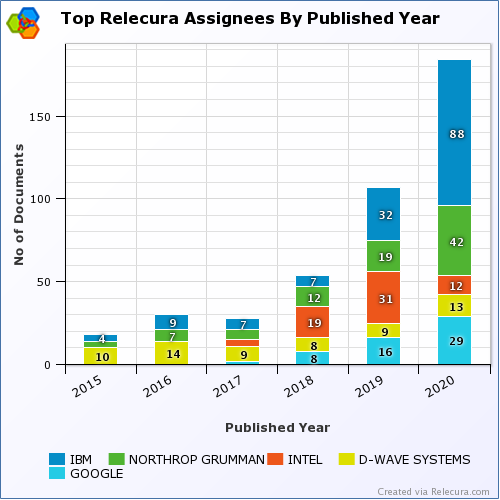}
	\caption{Top Assignees in Superconducting  Devices}
\end{figure}
\FloatBarrier
Superconductivity is the phenomenon of vanishing electrical resistance in some materials at sufficiently low temperatures characteristic of the material. It is a purely quantum phenomenon finding application in everything from strong electromagnets in MRI machines to qubits.

Superconducting qubits are the most sought after type of qubits in the industry; both IBM and D-Wave use them on their quantum computers. Even though the alternatives are microscopic systems like the electron, nucleons, etc., which are well isolated from the external noise, superconducting qubit, being a macroscopic system, is easy to couple with other qubits (\cite{superqubit}). All of the top five assignees are here for that reason.

\subsection{Quantum Sensing}
\begin{figure}[H]
	\centering
	\includegraphics[width=0.9\linewidth]{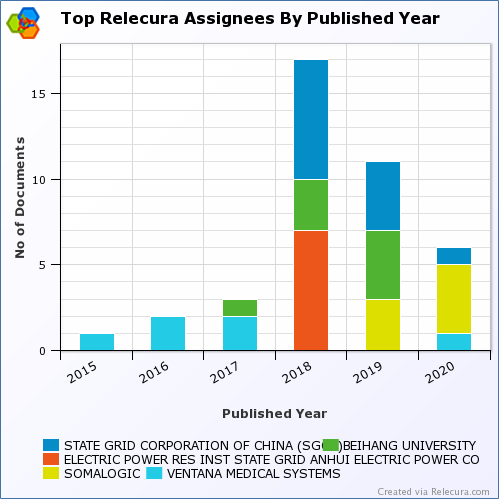}
	\caption{Top Assignees in Quantum  Sensing}
\end{figure}
\FloatBarrier

Quantum sensing capitalizes on the extreme sensitivity of quantum systems to external perturbations to make high precision sensors (\cite{sensing}). Though the atomic clock is an early example, present research goes hand in hand with quantum computers, with qubits often used as sensors to probe their quantum environment. Quantum dots can also be employed for sensing purposes for reasons discussed in Section \ref{qd}.

\section{Conclusions}

Quantum mechanics was one of the greatest revolutions in physics in the last century by any means. It explained so much of nature, gave high precision results, and drove the geniuses of the day nuts. Einstein's famous (correctly attributed) quote, "God does not play dice," elucidates his loathing of this theory. Physicists eventually learned to live with it because it is our best theory of nature. P.A.M. Dirac had the opinion that only such a theory that limits the scales at which we can meaningfully look can become a fundamental theory because otherwise, it will be an infinite ladder down (\cite{dirac}).  But quantum mechanics simplified some fields of physics like statistical mechanics, and the underlying mathematical richness of the theory has some elegance to it.

A similar benediction happened in the tech world from the onset of quantum technologies. We have turned every weirdness of quantum mechanics into something useful. The storage demand of quantum simulations scaled exponentially with the degrees of freedom, and we used it to our advantage in quantum computing. Quantum systems are highly sensitive to their environments, a huge challenge to overcome in the scaling of quantum computers, but that made them excellent candidates for enhanced sensors. The no-cloning theorem invalidated redundancy measures for error correction in quantum computing, but it is the cornerstone of quantum cryptography. The whole field encompasses the fruits of hard work and perseverance.

Quantum technologies are being launched into the industry at the right time for our civilization and earth. Nanotechnology and electronics are working in conjunction to put out devices that offer better performance and energy efficiency. The onset of better lighting solutions, semiconductor devices, etc., will lighten our energy budget, giving some cushioning in the transition towards renewable sources.  If ever realized quantum computers could do more than cutting down the energy we spend on computation. We can use quantum simulators to study molecular interactions and develop efficient industrial catalysts. Through full chemistry solutions, efficient alternatives to the Haber-Bosch process, responsible for 1\% global emissions, can be studied, and better battery chemistries can be invented. Quantum optimization algorithms can help us make better use of our resources to accommodate the growing population and climate change and someday even plan space missions.

While nanotechnology and quantum sensing materialized their promises in the form of products in the market, we should take quantum computing and communication with a grain of salt. The current state of the art, Noisy Intermediate-Scale Quantum era, is most likely to benefit quantum physics simulations at first (\cite{QS_2}), where decoherence errors are not undesirable, and maybe optimization with the success of companies like D-Wave. Many physicists think that quantum computing is decades away from making a significant contribution. Researchers are also exploring alternative computing paradigms like neuromorphic computing (\cite{neuromorphiccomputing}) and molecular computing(\cite{molecularcomputing}). Quantum computing is a passion project for most assignees in the field, who often have huge payrolls. Cryptographic systems that are unconditionally secure in all computation models, if used correctly, are already available (\cite{otp}), albeit at the expense of increased computational load. We could not find any replicable business models in quantum information; investments in R\&D in these fields are likely to show returns only in the long run.

There are no sure or impossible things in the tech world; there are only novel technologies and opportunity costs. Quantum technologies is a giant by its promises, attracting investments more than ever. IBM's Watson was introduced 64 years after the ENIAC, and humans went to the moon 66 years after the first flight. Quantum technologies have already put products on the market, and as a whole, the field is iteratively improving. Government agencies and industries are globally collaborating on an unprecedented scale to materialize the promises of this field. Many of today's billion-dollar businesses like the internet, nuclear energy, GPS, etc., were publicly funded or developed for military applications before there was a commercial market to them.
If anything, quantum technology is a prodigy, optimistically set to change everyone's life positively in the next decade at the latest.

\section{Acknowledgements}
This report is the culmination of my internship at Relecura Pvt. Ltd. (July-November 2020), Bengaluru, India, investigating patenting trends in quantum technologies and taxonomizing the patents. I thank the Relecura team, particularly Dr. George Koomullil, Dr. Hariharan Ramasangu, and Rohith Singh, for helping me understand the dynamics of the patent world using Relecura's AI-based tools and the invaluable insights they shared during the drafting of this report. Please reach out to Relecura (\href{mailto:support@relecura.com}{support@relecura.com}) for more information about the taxonomy and the tools used.

\bibliographystyle{plainnat}

\bibliography{ref}

\end{document}